# Quenching, plasmonic and radiative decays in nanogap emitting devices

Rémi Faggiani, Jianji Yang[#], and Philippe Lalanne[*]

Laboratoire Photonique, Numérique et Nanosciences (LP2N), UMR 5298, CNRS- IOGS-Univ. Bordeaux, 33400 Talence, France
[#]E-mail: jianji.yang.photonique@gmail.com
[*] E-mail: philippe.lalanne@institutoptique.fr

## Abstract

By placing a quantum emitter in the mouths of nanogaps consisting of two metal nanoparticles nearly into contact, significant increases in emission rate are obtained. This mechanism is central in the design of modern plasmonic nanoantennas. However, due to the lack of general knowledge on the balance between the different decay rates in nanogaps (emission, quenching, and metal absorption), the design of light-emitting devices based on nanogaps is performed in a rather hazardous fashion; general intuitive recipes do not presently exist. With accurate and simple closed-form expressions for the quenching rate and the decay rate into gap plasmons, we provide a comprehensive analysis of nanogap light emitting devices in the limit of small gap thickness. We disclose that the total decay rate in gap plasmons can largely overcome quenching for specifically selected metallic and insulator materials, regardless of the gap size. To confront these theoretical predictions, we provide a comprehensive numerical analysis of nanocube-type antennas in the limit of small gap thickness and further provide upper bounds for the photon-radiation efficiency.

## Introduction

Dielectric nanogaps in metals have unprecedented ability to concentrate light into deep-subwavelength volumes, which has propelled their use in a vast array of nanotechnologies and research topics (*1, 2*). For instance biological species can be manipulated at very low input powers by the field-gradients at nanogaps (*3*), photodetector with sizes well below the diffraction limit may be implemented with very fast response time (*4*), magnetic resonance supported by nanogap resonators can be utilized to realize negative-index metamaterials (*5, 6*), and feedgaps can be used to enhance the efficiency of high-harmonic nonlinear optical processes (*7*). The strongly confined field also profoundly alters light emission of quantum emitters placed in the nanogap by increasing optical excitation rates, modifying radiative and non-radiative decay rates, and altering emission directionality, leading to a new generation of ultra-compact nanoantenna architectures, such as bowties, spiral antenna, phased-array antenna (*8*), and to new applications for light-emitting devices (*9*), broadband single-photon sources (*10-12*), single-plasmon sources (*13, 14*), spasers or low-threshold nanolasers (*15-17*).

It is generally accepted that sub-wavelength architectures improve the exchange of optical energy with matter by strongly increasing spontaneous emission rates. As shown by numerical calculations and experimental measurements, it is also accepted that this enhancement can be achieved with relatively good efficiencies, ∼ 10-60% depending on the architecture. However the precise physical mechanisms that drive the emission of quantum emitters placed very close to metal surfaces in tiny gaps are not well understood. In particular, it is unclear from the literature why good efficiencies are achieved despite the proximity to the metal, why quenching is not the dominant decay channel, what is the ultimate efficiency, and whether this limit is impacted by the gap thickness or other parameters.

To further explore how optical antennas may lead to new regimes of light–matter interactions, it is important to first understand the different channel decays at play when quantum emitters in 2D nanogaps emit light in the immediate vicinity of metal surfaces and then draw a relationship between this basic situation and more complex problems of light emission and coupling with nanogap antenna architectures.

This is exactly the approach that is adopted in the present work. First, we provide a comprehensive analysis of the decay rates of quantum emitters placed in 2D planar nanogaps. So far, this has been discussed only with scattered numerical calculations performed for specific gap thicknesses and metal dielectric constants (*11, 18, 19*). In contrast, we derive a closed-form formula for the branching ratio between quenching and gap plasmon decays in the limit of small gap thicknesses, and then clarify the key material and geometrical parameters that drive the ratio. Counterintuitively, we evidence that the key parameters are the material permittivities, and not the gap thickness and that the gap plasmon decay surpasses the quenching decay for nanogaps fed with high-refractive-index materials and molecules polarized perpendicular to the gap interfaces.

Then we use the understanding gained from 2D planar structures to infer general recipes for designing efficient nanogap emitting devices. For that purpose, we model nanogap devices as gap-plasmon Fabry-Perot resonators and propose a phenomenological classification of nanogap antenna architectures, based on the tradeoff between quenching, absorption and free-space radiation rates, which may be monitored by controlling the gap-facet reflectance.

To set the classification against real structures, we analyze nanogap emitting devices formed by tiny nanocubes laying on metal surfaces. By scaling down the gap thickness and the cube dimensions to keep resonance in the visible, distinct behaviors within the same architecture family are comprehensively reviewed. Our analysis corrects inaccurate numerical results of the recent literature (*10, 12*), and definitely sets nanogaps with engineered facets as a very promising technological platform for light-emitting devices.

As a final point, we summarize the main results and provide a concluding discussion.

## Non-radiative decays in dielectric nanogaps

To start the analysis, let us consider the emission of a vertically-polarized molecule (treated as an electric dipole) buried in the middle of a polymer layer of thickness $2d$ of a metal-insulator-metal (MIM) planar stack. Two channels are available for the decay. Either the molecule excites the gap plasmon modes of the planar stack or quenches by directly heating the metal. We denote by $\gamma_{GSP}$ and $\gamma_{quench}$ the corresponding normalized decay rates (all decay rates are normalized by the vacuum decay rate hereafter). Figure 1 summarizes the main trends for the decay rates for an emission wavelength at 650 nm and silver nanogaps. First we find that the direct decay in the metal, $\gamma_{quench}$, scales as $d^{-3}$ as $d \to 0$. This scaling is understood from the local and static nature of quenching. Intuitively the quenching rate in a nanogap is expected to be ~ 2 times larger than the quenching rate $\gamma_{quench}^{SI}$ of the same vertical dipole on a *single interface* at the same separation distance *d*, and since

$$\gamma_{quench}^{SI} = \frac{3}{8\varepsilon_d (k_0 d)^3} \text{Im}\left(\frac{\varepsilon_m - \varepsilon_d}{\varepsilon_m + \varepsilon_d}\right)$$ as $d \to 0$ (19), the cubic scaling is well anticipated. In fact, $\gamma_{quench} \approx 2\gamma_{quench}^{SI}$ holds for $d > 10$ nm only; as smaller $d$'s, we found empirically with numerical calculations performed at 650 nm that the quenching rate $\gamma_{quench}$ is ~3 times larger than $\gamma_{quench}^{SI}$.

Importantly, we also find that the normalized decay rate into the gap plasmon modes of the planar stack also scales as the cube of the separation distance $2d$ between the metal films. What happens is the group velocity $v_g$ of gap plasmons drastically decreases as $d$ vanishes (20). Slowdown results in a strong field enhancement and the coupling to gap plasmons is boosted. Intuitively the fact that the plasmon decay rate and quenching have identical $d^{-3}$ scaling can be understood by considering that as $d \to 0$, gap plasmons completely lose their photonic character. They are highly damped and become of the same nature as the quenching field. By using a complex continuation approach to calculate the Green-tensor (21) and by assuming that the transverse electric and magnetic field components of the gap plasmon bear a flat profile within the gap, we have derived an analytical expression for $\gamma_{GSP}$ for very small gap thickness, $\gamma_{GSP} \approx \frac{12\pi\varepsilon_d}{(2k_0 d)^3 |\varepsilon_m|^2}$ (see Supporting Information for details), which evidences the inverse-cubic scaling.

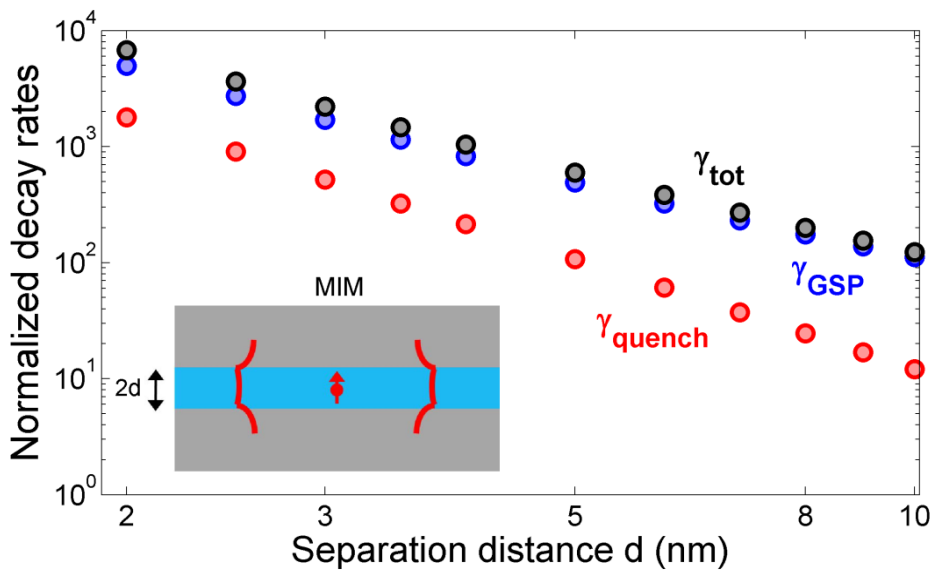

**Figure 1. Competition between several decay channels** for a vertical electric dipole emitting in the center of an Ag/polymer/Ag gap

of width 2*d*. The calculations are performed for an emission wavelength $\lambda_0$ = 650 nm. The refractive index of polymer is *n* = 1.4 and the silver permittivity is $\varepsilon_{Ag}$ = -17 + 1.15i (22). The decay rates into all channels ($\gamma_{tot}$) and gap plasmons ($\gamma_{GSP}$) and the quenching rate ($\gamma_{quench}$) are shown with black, blue, and red circles, respectively. All the decay rates are normalized by the decay rate in the vacuum.

Thus we obtain an analytical expression for an important figure of merit of planar nanogaps with vanishing gap thicknesses, namely the branching ratio *F* between gap-plasmon decay rates and quenching rates

$$F \approx \frac{\gamma_{GSP}}{3\gamma_{quench}^{SI}} = \frac{2\varepsilon_d}{Im(\varepsilon_m(\omega))}\left|\frac{\varepsilon_m(\omega)+\varepsilon_d}{\varepsilon_m(\omega)}\right|^2, \qquad (1)$$

with $\varepsilon_d$ and $\varepsilon_m$ denoting the dielectric and metal permittivities. The formula carries important hints:

- First, the ratio is independent of *d* for $k_0 d \to 0$, the first correction term being of order $O(k_0 d)^2$, and takes a universal expression that only depends on the dielectric constants.
- Second, for good metals, $\varepsilon_d << |\varepsilon_m(\omega)|$, one should bury the quantum emitter in a high-index material to enhance the branching ratio in the near- and far-infrared spectral regions.
- Third, the ratio solely depends on the losses encountered in polarizing the material, i.e. on $Im(\varepsilon_m(\omega))$, and not on the usual quality factor $\frac{-Re(\varepsilon_m)}{Im(\varepsilon_m)}$ of plasmonic materials, which gives an incontestable advantage to silver at visible frequencies, in comparison with gold or aluminum for instance.

Figure 2 shows typical branching ratios that can be obtained in the visible and near-infrared spectral regions with different metals and gap materials. The usefulness of the simple formula in Eq. (1) is reinforced by its ability to provide quantitative predictions, as evidenced with the comparison with fully-vectorial computational data (shown with marks) obtained for planar nanogaps with small gap sizes ($d$ = 2 nm). We emphasize that Eq. (1) is obtained in the asymptotic limit $k_0 d \to 0$; as the gap thickness increases beyond the quasi-static approximation, the ratio $F$ increases because quenching rapidly vanishes, and in this sense Eq. (1) actually sets a lower bound for the branching ratio. On overall, Eq. (1) represents a good compromise between simplicity or intuition and accuracy.

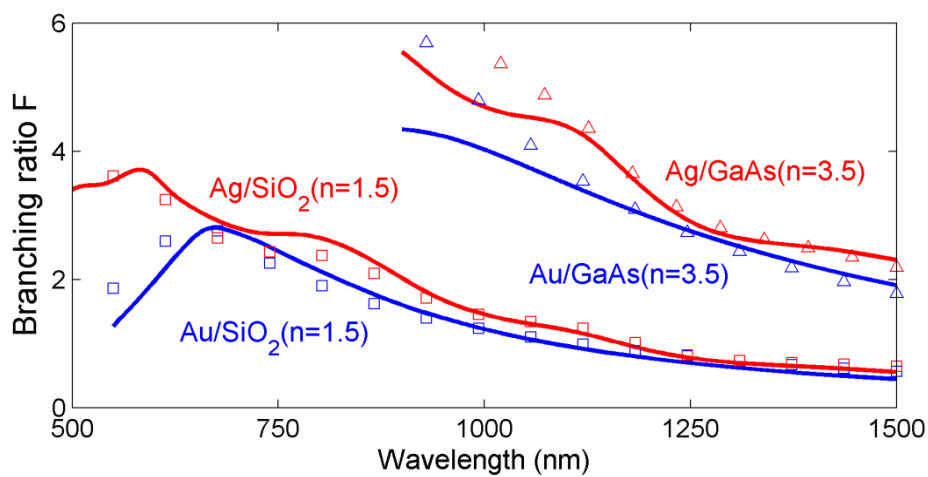

**Figure 2. Branching ratios for nanogaps formed with various materials at visible and near-infrared frequencies.** Fully-vectorial calculations (for $d$ = 2 nm) and analytical predictions from Eq. (1) are shown with markers and solid curves. Calculations made with Ag and Au are represented by red and blue colors, for dielectric (squares) and semiconductor (triangles) gap materials. Metal permittivities are taken from tabulated values (*22*).

In the fully-vectorial results shown in Figs. 1 and 2, quenching rates are found as the difference between the total decay rates (calculated as the Poynting-vector flux on a close surface surrounding the emitter) and the decay rates into gap plasmon modes (calculated using an open-source software (*23*)). Details on the calculation technique are provided in the Supporting Information along with a verification that the indirect derivation of the quenched energy actually corresponds to the absorption in the near-field zone ($< 0.02 \lambda_0^2$) of the emitter.

## Classification of nanogap emission devices

Decay into plasmon modes is often considered as detrimental, just like quenching. However it is of different nature since plasmons are coherent oscillations that may be transformed into free-space photons by scattering. This transformation is at the art of nanogap-device design. Intuitively, gap devices can be thought as Fabry-Perot nanoresonators with gap plasmon modes that bounce back and forth between two facets (*20, 24, 25*). The nanoresonator modes can couple to different decay channels, i.e. to free-space photons and surface plasmon polaritons (SPPs) for nanoresonators surrounded by metal films, with normalized rates $\gamma_{rad}$ and $\gamma_{SP}$ respectively. They also give rise to a new non-radiative decay, the mode absorption, with a decay rate $\gamma_{abs}$. Unlike quenching which is intrinsically determined by proximity to metal, $\gamma_{abs}$ is determined by the nanoantenna design and particularly the reflectivity *R* of the gap facets. Thus, from the sole knowledge acquired on the 2D planar structures and according to values of R that govern the resonance strength, we may distinguish three different nanogap-device categories, as illustrated in the classification of Fig. 3.

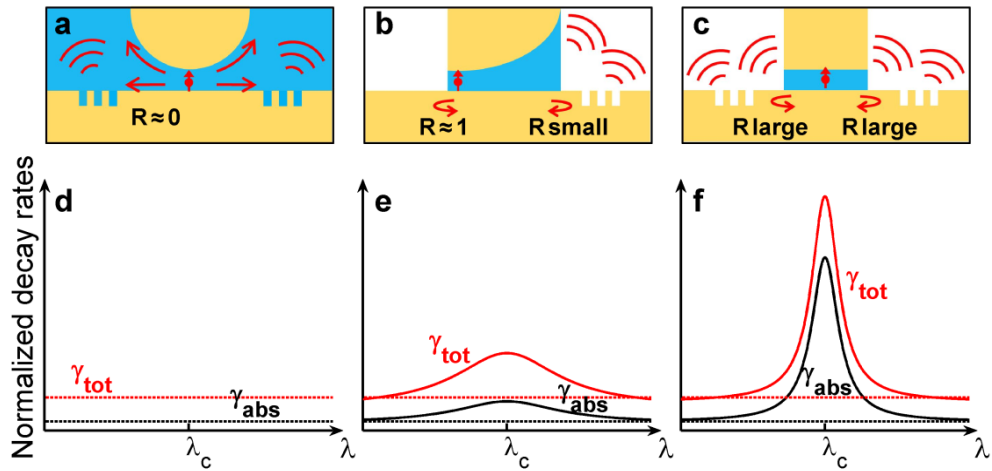

**Figure 3. Classification of planar nanogap emitting devices with different degrees of decay rate enhancements. (a)** Tapered devices with large photon-radiation efficiencies. Dipole emission is initially captured by gap plasmons and then adiabatically (low reflection R~0) converted into photons. Grooves etched in the metal film help conversion of launched SPPs into free-space photons. **(b)** Nanoresonators with controlled facet reflectivities. **(c)** Nanoresonators with strong facet reflectivities. **(d)**-**(f)** Corresponding decay rates. **(d)** Due to the tapering, $\gamma_{tot} \approx \gamma_{GSP} + \gamma_{quench}$, where $\gamma_{GSP}$ and $\gamma_{quench}$ are the gap-plasmon and quenching decay rates

obtained in a planar nanogap with a thickness equal to the mouth thickness. (**e**) The weak reflection in (**b**) results in a broadband rate enhancement with a large photon-radiation efficiency. (**f**) The strong reflection in (**c**) results in a narrowband Fabry-Perot resonance; $\gamma_{tot}$ can be considerably boosted, but the non-radiative decay $\gamma_{abs}$ due to cavity absorption lowers the photon-radiation efficiency.

Almost nil reflectivity is implemented in tapered nanogaps formed for instance by curved and flat metal surfaces (Fig. 3a) by adiabatically converting the slow gap-plasmons generated at the mouth (*26-28*) into free-space photons and SPPs launched on the flat metal surface surrounding the device. The SPPs can be further converted into photons with groove arrays for instance. Thus the total decay rate is expected to be equal to $\gamma_{tot} \approx \gamma_{GSP} + \gamma_{quench}$, where $\gamma_{GSP}$ and $\gamma_{quench}$ are the gap-plasmon and quenching decay rates obtained in a planar nanogap with a thickness equal to the mouth thickness. For full SPP conversion into free-space photons, the photon-radiation efficiency is thus limited by quenching and is bounded by $F/(F+1)$, a value that can be as large as 0.75 at an emission wavelength of 600 nm for Ag/polymer gaps (see Fig. 2).

By contrast, for strong reflectivity favored by large impedance mismatch at the facet of tiny gaps (*10, 16*), see Fig. 3c, the total decay rate is considerably boosted and quenching becomes completely negligible. However, the photon-radiation efficiency is limited by the absorption of gap plasmons in the tiny gap; it is expected to be much smaller than the value reached for adiabatically-tapered antennas.

Figure 3b illustrates a promising class of nanogap antenna, with intermediate values of the facet reflectivity. Spontaneous-decay rates larger than $\gamma_{GSP}$ are implemented, but the photon-radiation efficiency that is limited by quenching and absorption in the cavity may approach or even exceed the upper bound value of the *R* ≈ 0 case.

**Nanocube geometry**

After this qualitative discussion, it is interesting to set the classification of Fig. 3 against real nanogap emitting-device technologies. For that purpose, we consider devices formed by a tiny dielectric layer sandwiched between a metallic nanocube and a metallic substrate. This geometry that has been recently studied (*10, 12*) is particularly suitable for exploring the three categories of the classification, since by scaling down the cube dimension, a whole family of gap emitting devices with distinct facet reflectivities can be straightforwardly designed and studied at the same resonant visible wavelength. But first let us begin by explaining how we calculate the photon-radiation efficiency and estimate the respective impact of quenching and gap-plasmon absorption. This is necessary because our results differ from those reported in (*10, 12*), at least by a factor two for the efficiency, and it is important to understand the reason.

We consider the same nanocube geometry as in (*10*), with a vertically-polarized emitting dipole. The latter is placed in the middle of an 8-nm-thick polymer gap at a cube corner, where maximum coupling with the resonance mode is achieved. Using COMSOL multiphysics, we first calculate the normalized total decay rate $\gamma_{tot}$ by integrating the total power radiated around the source. Consistently with (*10*), at resonance $\gamma_{tot}$ (black circles) is as large as $10^4$, a value which represents a tenfold enhancement, compared to the normalized gap-plasmon decay rate $\gamma_{GSP}$ (cyan circles) obtained for a planar nanogap with the same materials and gap thickness.

To provide a deeper insight, we also calculate the normalized decay rate $\gamma_{mode}$ (i.e. the Purcell factor) into the fundamental magnetic-dipolar nanocube mode (*12*). $\gamma_{mode}$ is calculated by using a resonance-mode theory recently developed (*29, 30*) to the analysis of plasmonic nanoresonators. Owing to the very small mode volume $V = (84,000+8,000i)$ nm$^3$ of the magnetic-dipolar mode, we find that 95% of the total decay is actually funneled into the resonance mode at resonance. The quenching rate $\gamma_{quench}$ (red circles) is then calculated as $\gamma_{quench} = \gamma_{tot} - \gamma_{mode} - \gamma_{quad}$, where $\gamma_{quad}$ represents a residual decay into a quadrupolar mode that resonates in the green (*31*). Unlike intuitive statements in (*10*), our calculations indicate that quenching (red circles) is playing a negligible role, as its rate only represents ~ 2% of $\gamma_{tot}$ at resonance. Then, using an open source code that computes the radiation diagrams of free-space and guided waves (*23*), we calculate the normalized decays into free space photons and SPPs launched around the nanocube,

$\gamma_{rad}$ and $\gamma_{SP}$, respectively. We find that ~60% of the mode energy is dissipated into heat, and the remaining 40% equally decays into free-space photons (20%) and SPPs (20%) that are launched on the flat metal interface surrounding the cube. This suggests the great potential of nanocubes for implementing plasmon sources. The present prediction $\gamma_{rad}/\gamma_{tot} \approx 20\%$ differs from the 50% photon-radiation efficiency calculated in (*10, 12*) for the same geometrical parameters. We believe that the discrepancy is due to the fact that in (*10, 12*), the photon-radiation efficiency is inferred from a direct computation of the Poynting-vector flux on a close surface surrounding the nanocube, without separating the respective contributions of the radiated photons and surface plasmons with a near-to-far-field transform.

Clearly, the nanocube antenna with a facet reflectivity of ~0.85, a value deduced from results reported in (*20*), belongs to the category of nanogap antennas with large facet reflectivities, i.e. case (c) in the classification of Fig. 3.

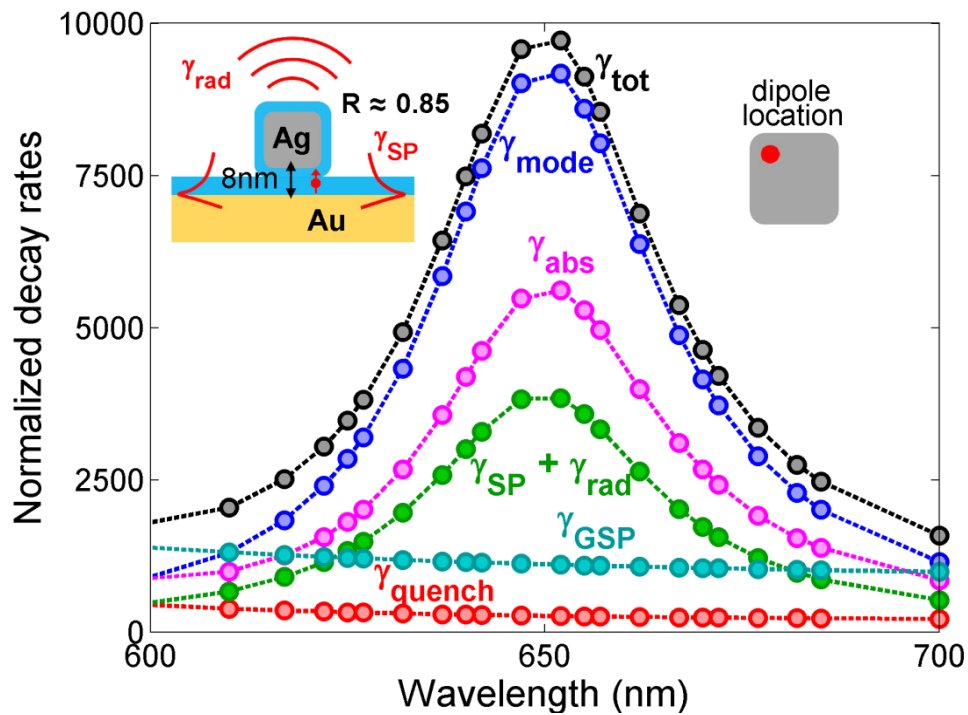

**Figure 4. The decay channels of nanocube antennas.** Calculated normalized decay rates of a vertical electric dipole at the corner (insets) of the antenna as a function of the wavelength. $\gamma_{tot}$, $\gamma_{mode}$ (decay into the fundamental magnetic mode), $\gamma_{abs}$ (antenna absorption), $\gamma_{rad} + \gamma_{SP}$ (sum of the radiative and SPP decay rates)

and $\gamma_{quench}$ are shown by black, blue, magenta, green, and red circles, respectively. $\gamma_{GSP}$ (cyan) represents the decay into the gap plasmon of a planar nanogap of the same materials and gap thickness. Left inset: cross-sectional view of the nanoantenna: a silver nanocube (side length 65 nm) with a 3-nm polymer coating (n=1.4) is placed on a gold substrate covered by a 5-nm polymer (8-nm gap). The molecule is represented as a red arrow placed in the middle of gap. Right inset: top view of the cube showing the position of the dipole. The frequency-dependent permittivities, $\varepsilon_{Ag}$ and $\varepsilon_{Au}$, of silver and gold are taken from tabulated data (*22*), $\varepsilon_{Ag} = -17 + 1.15i$ and $\varepsilon_{Au} = -9.7 + 1.04i$ @ $\lambda$ = 650 nm.

We are now ready to study the nanocube performance for various thicknesses. For that, at every thickness, we adapt the cube size to maintain the magnetic-dipolar resonance at $\lambda_0$ = 650 nm and repeat the previous modal analysis. The results for total decay rate $\gamma_{tot}$ and the external efficiency $(\gamma_{rad} + \gamma_{SP})/\gamma_{tot}$ defined as the normalized decay into SPPs and photons are displayed in Fig. 5a as a function of the gap thickness. The latter is shown to importantly impact the nanocube performance. As the thickness reduces, the facet reflectivity increases (*20*) and accordingly, $\gamma_{tot}$ strongly increases. However, the coupling to outgoing channels also decreases and the enhancement of the total decay rates is accompanied by a sudden drop of the external efficiency, from 80% for *d* = 20 nm to 15% for *d* = 4 nm.

For the sake of comparison, in Fig. 5b, we display $\gamma_{tot}$ and the external efficiency $(\gamma_{rad} + \gamma_{SP})/\gamma_{tot}$ (again, $\gamma_{SP}$ denotes the decay to the SPPs on the flat metal surface, not gap plasmons) for a perfectly-tapered antenna (case (a) of the classification in Fig. 3). The predictions are obtained from planar nanogap calculations only, by assuming that $(\gamma_{rad} + \gamma_{SP})$ is equal to the gap-plasmon decay rate $\gamma_{GSP}$ in a planar nanogap with a thickness equal to the mouth thickness. This amounts to assume that gap plasmons are fully converted by the tapering structure into SPPs and/or photons and this provides an upper bound for the photon-radiation efficiency. As shown by a comparison between Figs. 5a and 5b, smaller decay rates are achieved with the tapered antenna, but significantly larger external efficiencies are also obtained. Impressively, we predict that large efficiencies >70% with large normalized emission rates ~$10^3$ are achieved for tiny nanogaps. Even if this prediction is optimistic, it opens important perspectives for spontaneous light emission in general, and definitely sets nanogaps as a relevant technological platform. The future success of the platform will depend on fabrication and material issues, and on our ability to engineer facet reflectivities adequately.

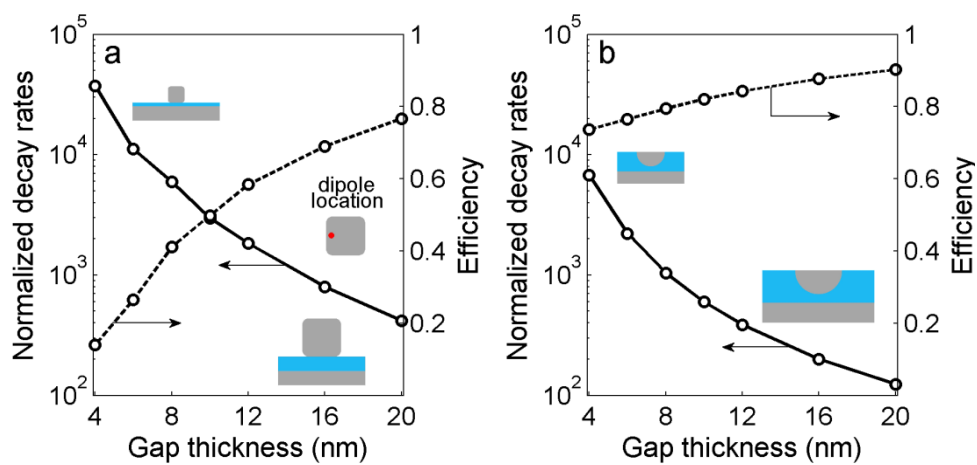

**Figure 5. Patch versus tapered-nanogap antennas with polymer gaps. (a)** Patch antennas. The calculations are performed for a vertical electric dipole located close to the center of the antenna facet (inset). The cube size varies with the gap thickness to maintain the resonance wavelength at 650 nm (side lengths are 47, 56, 65, 70, 75, 80 and 85 nm, for gap thickness $d$ = 4, 6, 8, 10, 12, 16 and 20 nm). (**b**) Perfectly tapered-nanogap antennas. The performances are predicted with planar-nanogap calculations by assuming perfect tapering of the gap-plasmons into SPPs and/or radiative photons. In (**a**) and (**b**), the normalized total decay rates $\gamma_{tot}$ are shown with solid curves and the external efficiencies $(\gamma_{rad} + \gamma_{SP})/\gamma_{tot}$ with dashed curves.

## Concluding discussion

Emitting photonic devices with quantum emitters embedded in nanogaps for operation at visible and near-infrared frequencies can provide large spontaneous emission rate enhancements and good photon-radiation efficiencies, because the decay into slow gap-plasmons is considerably boosted, and quenching is thus effectively overcome. This is particularly true for gaps with high-refractive-index insulators sandwiched between good metals, since the branching ratio $\propto \varepsilon_d / Im(\varepsilon_m(\omega))$ between gap-plasmon decay rates and quenching rates reach values as large as 80% for semiconductor gaps operated at near-infrared frequencies. The dominant character of plasmonic decays for small gaps has a direct impact on the design and performance of nanogap emitting devices. First, the high decay rates found in planar nanogaps can be harnessed to realize tapered antennas (Fig. 3a) offering strong decay rate enhancements ($\approx 10^2$–$10^3$) and large photon-radiation efficiencies limited by the branching ratio between gap-plasmon decay rates and quenching rates, see Eq. (1). Second, even larger rate enhancements can be even achieved in nanogap cavities (Fig. 3c), which exhibit strong resonances owing to strong reflection of gap plasmons at the cavity facets. In return, the photon-radiation efficiency is significantly reduced by the cavity absorption, as indicated by the analysis of the state-of-the-art nanocube devices. Third, a better balance between decay rate enhancement and photon-radiation efficiency may be reached with nanogap antennas with engineered facet reflectivities (Fig. 3b) for which a delicate engineering of the facets and a precise choice of the gap and metal materials may lead to acceleration decay rates greater than $10^3$ with significant photon-radiation efficiency, $\approx 50\%$.

Certainly, the main strength of nanogap light emitting devices is the capability to boost the spontaneous emission rate over a broad bandwidth with potentially easy electrical contacting. This might be useful for increasing quantum yield. However, the present analysis seems to indicate that it will be hard to achieve extremely high efficiencies ($\approx 1$), as required for some quantum-information protocols, and that the branching of Eq. 3 appears as a barrier for the photon-radiation efficiency which will be hard to overcome.

ASSOCIATED CONTENT
**Supporting Information**.
In the supporting information, some technical details are provided.
(1) Indirect calculation of quenching. (2) Verification of the indirect quenching calculation and localized nature of the quenched fields. (3) Asymptotic expression of $\gamma_{GSP}$ for MIM stacks with vanishing gap thickness. This material is available free of charge via the Internet at http://pubs.acs.org.

AUTHOR INFORMATION
**Corresponding Authors**

#E-mail: jianji.yang.photonique@gmail.com
*E-mail: philippe.lalanne@institutoptique.fr
**Present Address**
#Department of Electrical Engineering, Stanford University, Stanford, California 94305, USA

**Notes**
The authors declare no competing financial interest.

**Acknowledgment.** The authors thank Jean-Paul Hugonin for computational helps. R. F. acknowledges financial support from the French "Direction Générale de l'Armement" (DGA).

# Supporting Information of
# Quenching, plasmonic and radiative decays in nanogap emitting devices
Rémi Faggiani, Jianji Yang#, and Philippe Lalanne*

Laboratoire Photonique, Numérique et Nanosciences (LP2N), UMR 5298,
CNRS-IOGS-Univ. Bordeaux, 33400 Talence, France
# E-mail: jianji.yang.photonique@gmail.com
* E-mail: philippe.lalanne@institutoptique.fr

## Content

## 1. Indirect calculation of quenching

For the dipole emission into an MIM stack problem shown in Fig. 1 in the main text, the dipole decays either into gap plasmons modes ($\gamma_{GSP}$) or couples directly to the metal (quenching, $\gamma_{quench}$). To estimate the quenching, we first calculate the total decay rate by integrating the total power emitted by the source over a box surrounding the source. Then with an open-source near-to-far-field transform (NFFT) tool (*1*), we obtain the decay into the propagative modes ($\gamma_{GSP}$). Finally $\gamma_{quench}$ is *indirectly* calculated as the difference $\gamma_{quench} = \gamma_{tot} - \gamma_{GSP}$.

Note that, in previous works on dipole emission in MIM stacks (*2, 3*), $\gamma_{quench}$ is directly calculated by integrating the power coupled into all evanescent waves, i.e. into decay channels that are not associated to propagative modes. In contrast, in our work $\gamma_{quench}$ is obtained in an indirect way, mainly because the calculations of $\gamma_{tot}$ and the NFFT implementation are accurate and efficient. The validity of the indirect calculation by difference is shown in the next section.

## 2. Verification of the indirect quenching calculation and localized nature of quenched fields

The following verification is motivated by two main reasons:
1/Power does not sum up in absorbing media, and thus strictly speaking, we cannot decompose the total decay rate as a sum of decays into different channels. This decomposition is valid only in the limit of "weak absorptions". Nothing guaranties that.
2/Taking the example of metal/dielectric interfaces for simplicity, it is well known that absorption in the metal has (at least) three distinct origins: SPP launching, a quasi-cylindrical wave launching that SPP alters the absorption over an area of ≈ $100\lambda^2$ around the source (4), and a localized absorption localized in the near-field of the source. This again implies that quenching (defined as the localized absorption) cannot be strictly obtained as a difference between the total decay and the sum of the decays into propagative modes.

It is therefore important to check numerically that this different in power decays effectively corresponds to a localized absorption.

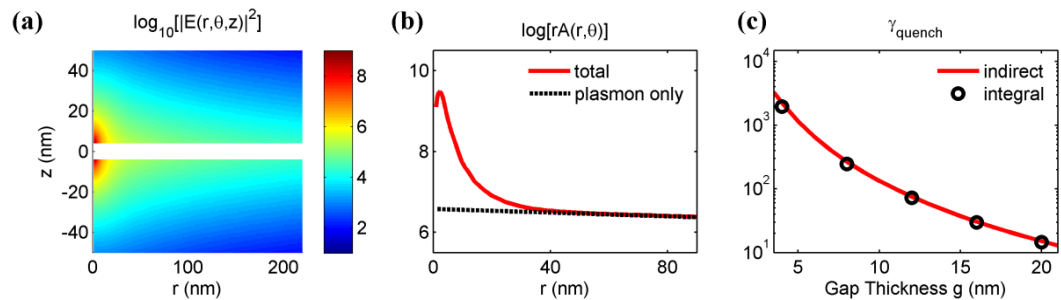

**Figure S1. Localized nature of quenching**. (**a**) Intensity of the field excited at a wavelength of 650 nm by a vertical electric dipole in the center ($z = 0$ and $r = 0$) of an Ag/Polymer/Ag MIM stack with a gap thickness of $g$ =8 nm. The field in the gap is not represented to better show the field in metal claddings.(**b**) Radial plots of $rA(r,\theta)$ as a function of $r$ (red solid) and of the exponential damping of gap plasmon (black dashed). (**c**) $\gamma_{quench}$ obtained by indirect calculation ($\gamma_{quench} = \gamma_{tot} - \gamma_{GSP}$, red line) and by integrating the density of quenched power (black circles), for varying gap thicknesses. In the calculation, the refractive index of polymer is $n = 1.4$ and the relative permittivity of silver is $\varepsilon_{Ag} = 17 + 1.15i$.

For that purpose, we consider a vertically-polarized dipole emitting at a wavelength of 650 nm in an MIM with a gap thickness $g = 8$ nm. Since the MIM stack laterally extends to infinity, all the emitted power is absorbed by the metal. The total field is shown in Fig. S1a. We can clearly see an intense field in the metal claddings in the immediate vicinity of the source. In a cylindrical coordinate $(r, \theta, z)$, we define the absorbed power density as $A(r,\theta) = 0.5\omega \int \text{Im}(\varepsilon)|\mathbf{E}(r,\theta)|^2 dz$. The total absorbed power (or total emitted power) $P_{tot}$ is simply $P_{tot} = \iint rA(r,\theta)drd\theta$. In Fig. S1b, the logarithm of $rA(r,\theta)$ is plotted with a red curve at an arbitrary azimuth (for a vertical dipole, the whole system is azimuthal-independent). The exponentially decaying tail (the linear part in the logarithm coordinates) of $rA(r,\theta)$ corresponds to the gap-plasmon damping $e^{-2\text{Im}(k_{GSP})r}$, with $k_{GSP}$ denoting the propagation constant. A backward extrapolation of the tail (black dashed) to $r = 0$ offers a clear distinction between the respective contributions of gap plasmons and quenching (which occurs in a very short length scale around the source) to total absorption. The quenching area (between the red and black curves) is at deep subwavelength scale, revealing the localized nature of quenching.

As shown in Fig. S1c, quenching rate $\gamma_{quench}$ (black circles) obtained by directly integrating the density of quenched power in real space matches $\gamma_{quench}$ obtained with the indirect calculation (red lines). The quantitative agreement validates the indirect calculation.

## 3. Asymptotic expression of $\gamma_{GSP}$ for MIM stacks with vanishing gap thickness

The objective of this section is to derive an analytical formula for the emission of a vertically polarized electric dipole source emitting an MIM stack. The following derivation relies on a formalism that is developed in (*1*).

### 3.1. Power coupled to gap plasmons in MIM stacks

For an electric current **J** that is vertically polarized (**J** = **J**$_z$, see Fig. S2a) and placed at $r = 0$, the excited gap plasmon field $\mathbf{E}_{GSP}(r,\theta,z)$ can be written as

$$\mathbf{E}_{GSP}(r,\theta,z) = c_{GSP}\hat{\mathbf{E}}^+_{GSP}(k_{GSP}\,r,z), \tag{S1}$$

where $c_{GSP}$ denotes the mode amplitude (a constant to be determined), $\hat{\mathbf{E}}^+_{GSP}(k_{GSP}\,r,z)$ describes the $r$- and $z$-dependent mode profile, and the superscript '+' indicates outgoing mode (propagating from $r=0$ to $r=\infty$).

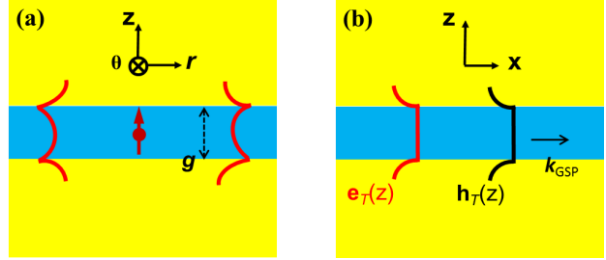

**Figure S2. Dipole emission in an MIM stacks. (a)** We consider the emission of a vertical electric dipole placed in the dielectric nanogap at $r=0$. **(b)** In an MIM stack with a narrow gap ($g\rightarrow 0$), the transverse electric and magnetic field-components can be assumed to be independent of the transverse coordinate.

According to the unconjugated form of Lorentz reciprocity, the plasmon amplitude can be found by a dot product of $\mathbf{J}_z$ and the ingoing mode $\hat{\mathbf{E}}^-_{GSP}(k_{GSP}\,r,z)$ as (*1*)

$$c_{GSP} = \hat{\mathbf{E}}^-_{GSP}\bullet\mathbf{J}_z/N_{GSP}. \tag{S2}$$

In Eq. (S2), $N_{GSP} = 16/k_{GSP}$ denotes the mode normalization coefficient (*1*) and the minus superscript '–' refers to ingoing modes (propagating from $r=\infty$ to $r=0$). Accordingly the total power carried by the gap plasmon is found as

$$P_{GSP} = 4|c_{GSP}|^2/|k_{GSP}|. \tag{S3}$$

According to Eqs. (S2) and (S3), we need the $z$-component electric field of the mode. The $z$-component of the gap plasmon can be written as $\hat{\mathbf{E}}^\pm_{GSP,z} = H^\pm_0(k_{GSP}\,r)e_T(z)$ (*1*), with $H^\pm_n$ denoting the Hankel function of the first ('+') or second ('–') kind of order $n$, $e_T(z)$ denoting the transverse electric field of the $z$-dependent mode profile (see Fig. S2b). Note that since the dipole source is placed at $r=0$, the Hankel functions diverge at $r=0$. Therefore a special trick to avoid singularity is applied (*1*), and Eq. (S3) is rewritten as

$$P_{GSP}(z) = |\mathbf{J}_z|^2|e_T(z)|^2|k_{GSP}|/16. \tag{S4}$$

### 3.2. Mode profile of gap plasmon in MIM stacks with very small gap

To use Eq. (S4) for deriving a closed-from expression, we need to calculate $e_T(z)$ and $k_{GSP}$. In principle, the mode profile can be calculated analytically (5); however, as we are solely interested by vanishing gaps ($g\to 0$), the mode profile can be found in a simple manner by assuming that the transverse electric $e_T(z)$ and magnetic $h_T(z)$ field components inside the gap are uniform (Fig. S2b). Thus we have

$$\begin{cases} e_T(z) = e_0, h_T(z) = h_0; & \text{for } |z| \leq g/2 \\ e_T(z) = e_0 \frac{\varepsilon_d}{\varepsilon_m} \exp(-\tau|z - g/2|), h_T(z) = h_0 \exp(-\tau|z - g/2|) & \text{for } |z| > g/2 \end{cases} \quad (S5)$$

where $e_0$ and $h_0$ denote the amplitudes of electric and magnetic components inside the gap, $\tau$ denotes the damping term of the plasmon in the metal cladding. Applying Eq. (S5) into unconjugated form of mode normalization based on Lorentz reciprocity (a normalization method valid for lossy modes) (6, 7) $\int e_T(z) h_T(z) dz = -2$, we find easily that

$$e_0 h_0 \left( g + \frac{\varepsilon_d}{\varepsilon_m} \frac{1}{\tau} \right) = -2. \quad (S6)$$

If $g \to 0$, $k_{GSP}$ can be expressed as $k_{GSP} = -\frac{2\varepsilon_d}{\varepsilon_m g}$ (8), with $\varepsilon_d$ and $\varepsilon_m$ denoting the gap (dielectric) and metal permittivities, and $\tau$ ($\tau^2 = k_{GSP}^2 - k_0^2 \varepsilon_m$) can be found as

$$\tau \approx -\frac{2\varepsilon_d}{\varepsilon_m g}. \quad (S7)$$

Applying Eq. (S7) into Eq. (S6), we obtain the important (and simple) relation

$$e_0 h_0 \approx -\frac{4}{g}. \quad (S8)$$

Then by applying $h_T(z) = \frac{\omega \varepsilon_0 \varepsilon_d}{k_{GSP}} e_T(z)$ into Eq. (S8), which is valid within the dielectric gap and can be easily found for any slab waveguide, we find

$$e_T(z) = e_0 = \sqrt{8/(\omega \varepsilon_0 \varepsilon_m g^2)} \text{ for } |z| \leq g/2. \quad (S9)$$

### 3.3. Asymptotic expression of $\gamma_{GSP}$ for very small gaps

Incorporating Eq. (S9) into Eq. (S4), we find

$$P_{GSP} = \frac{\varepsilon_d}{\omega g^3 \varepsilon_0 |\varepsilon_m|^2} |\mathbf{J}_z|^2 \quad (g \to 0). \tag{S10}$$

The radiated power of the same dipole in vacuum being $P_0 = \frac{\omega^2 n}{12\pi\varepsilon_0 c^3}|\mathbf{J}_z|^2$, we finally get the normalized decay rate as (*9*)

$$\gamma_{GSP} = \frac{P_{GSP}}{P_0} = \frac{12\pi\varepsilon_d}{(k_0 g)^3 |\varepsilon_m|^2}. \tag{S11}$$

Clearly, Eq. (S11) indicates that decay to gap plasmon varies as $d^{-3}$ (*d* denoting the dipole-metal distance). We recall that Eq. (S11) relies on the approximation that the transverse components of the Gap-plasmon mode does not vary with the transverse coordinate. This is all the more accurate as the gap thickness is small.